\documentclass[useAMS,usenatbib,usegraphicx,letterpaper]{mn2e}
\usepackage{amssymb}

\usepackage{multirow}
\usepackage{epsfig}
\usepackage{subfigure}

\def\mnras{MNRAS}
\def\apj{ApJ}
\def\apjs{ApJS}

\title[$M_{\rm BH}$ - $\sigma$ relation in FSRQs]
{$M_{\rm BH}$ - $\sigma$ relation in SDSS flat-spectrum radio
quasars}
\author[M. F. Gu, Z. Y. Chen \& X. Cao]
      {Minfeng Gu$^{1}$\thanks{gumf@shao.ac.cn},
        Zhaoyu Chen$^{1,2}$, Xinwu Cao$^{1}$
\\ \\
  $^1$ Key Laboratory for Research in Galaxies and Cosmology,
Shanghai Astronomical Observatory, Chinese Academy of Sciences, 80\\
Nandan Road, Shanghai 200030, China\\
  $^2$Graduate School of the Chinese Academy of Sciences, Beijing
100039, China}

\pagerange{\pageref{firstpage}--\pageref{lastpage}} \pubyear{2009}

\usepackage{times}

\begin{document}

\label{firstpage}

 \maketitle

\begin{abstract}
The relationship between the black hole mass and velocity dispersion
indicated with [O III] line width is investigated for a sample of 87
flat-spectrum radio quasars (FSRQs) selected from SDSS DR3 quasar
catalogue. We found the $M_{\rm bh} - \sigma_{\rm [O III]}$ relation
is deviated from Tremaine et al. relation for nearby inactive
galaxies, with a larger black hole mass at given velocity
dispersion. There is no strong evidence of cosmology evolution in
$M_{\rm bh} - \sigma_{\rm [O III]}$ relation up to $z\sim0.8$. A
significant correlation between the [O III] luminosity and Broad
Line Region (BLR) luminosity is found. When transferring the [O III]
luminosity to Narrow Line Region (NLR) luminosity, the BLR
luminosity is, on average, larger than NLR one by about one order of
magnitude. We found a strong correlation between the synchrotron
peak luminosity and NLR luminosity, which implies a tight relation
between the jet physics and accretion process.
\end{abstract}


\section{Introduction}

The evolution of black holes has been shown to be closely coupled to
that of their host galaxies for normal galaxies, mainly through the
tight correlation between the central black hole mass and the bulge
mass and luminosity (Kormendy \& Richstone 1995; Magorrian et al.
1998), and tighter one between the central black hole mass and
stellar velocity dispersion $\sigma_{*}$ in the galactic bulge
(Ferrarese \& Merritt 2000; Gebhardt et al. 2000a). The latter
relation has been established by Tremaine et al. (2002) for a sample
of 31 nearby inactive galaxies as
\begin{equation}
\rm log~\it (\frac{M_{\rm BH}}{\rm M_{\odot}})=\rm (8.13\pm0.06) +
(4.02\pm0.32)~ log ~ \it (\frac{\sigma_{*}}{\rm 200~ km~s^{-1}}).
\end{equation}
For some AGNs with available bulge velocity dispersion and the
reverberation mapping black hole mass, Gebhardt et al. (2000b) and
Ferrarese et al. (2001) found that these AGNs also follow the
$M_{\rm bh} - \sigma_{*}$ relation founded in the nearby inactive
galaxies. With the aim to constrain the nature and evolution of AGNs
and the advantage of relative higher redshift of AGNs, the $M_{\rm
bh} - \sigma_{*}$ relation and its evolution have been extensively
explored for different AGN populations, such as radio-quiet,
radio-loud AGNs, narrow line Seyfert 1 galaxies and young radio
galaxies et al. (e.g. Shields et al. 2003; Bian \& Zhao 2004;
Bonning et al. 2005; Liu \& Jiang 2006; Salviander et al. 2007; Bian
et al. 2008; Shen et al. 2008; Wu 2009a), however it is still in
large debates whether AGNs generally follow the $M_{\rm bh} -
\sigma_{*}$ relation.

While the derivation of $M_{\rm bh}$ from AGNs broad line widths and
continuum (and/or broad line) luminosity is now well established
(e.g. Kaspi et al. 2000; Greene \& Ho 2005; Kong et al. 2006), the
measurements of $\sigma_{*}$ are generally difficult for quasars,
due to the faintness of the host galaxy and the relative brightness
of active nucleus. When stellar velocities dispersion cannot be
measured, the line widths of the narrow emission lines (e.g. [O
III], [S II]) usually can be used to trace $\sigma_{*}$ (Nelson \&
Whittle 1996; Greene \& Ho 2005). Nelson \& Whittle (1996) compared
the bulge magnitudes, [O III] $\lambda$5007 line widths, and stellar
velocity dispersions in Seyfert galaxies, adopting $\sigma_{\rm [O
III]}=\rm FWHM ([O III])/2.35$, and found, on average, good
agreement between $\sigma_{\rm [O III]}$ and $\sigma_{*}$, although
$\sigma_{\rm [O III]}$ shows more scatter than $\sigma_{*}$ on a
Faber-Jackson plot. This implies that the kinematics of narrow-line
region (NLR) gas is largely dominated by the bulge gravitational
potential, therefore, can be effectively used as a substitute for
$\sigma_{*}$ of galaxy bulges. Using [O III] line width as
surrogates for $\sigma_{*}$, Nelson (2000) claimed that the $M_{\rm
bh} - \sigma_{*}$ relation for normal galaxies and active galactic
nuclei is preserved. Albeit some defect of [O III] line profile
(e.g. asymmetry and non-Gaussian) and other surrogates proposed
(e.g. [S II], Greene \& Ho 2005), the width of [O III] is still most
commonly used as surrogate for $\sigma_{*}$ in AGNs (e.g. Bonning et
al. 2005; Bian et al. 2008).

The formation of radio jets is still a unresolved issue in AGNs
research. While radio-quiet AGNs roughly follow the $M_{\rm bh} -
\sigma_{*}$ relation of normal galaxies, radio-loud ones deviate, in
the sense that radio-loud objects have, on average, larger black
hole mass than radio-quiet objects for a given velocity dispersion
(e.g. Bian \& Zhao 2004; Bonning et al. 2005; Liu \& Jiang 2006;
Salviander et al. 2007; Bian et al. 2008; Shen et al. 2008). Among
the population of radio-loud AGNs, FSRQs represent an extreme class,
which are generally characterized by strong and rapid variability,
high polarization, and apparent superluminal motion. These extreme
properties are generally interpreted as a consequence of non-thermal
emission from a relativistic jet oriented close to the line of
sight. In this letter, we investigate the relationship between the
black hole mass and $\rm \sigma_{\rm [O III]}$ for a sample of 87
SDSS FSRQs based on the spectral analysis of a larger FSRQs sample
in Chen et al. (2009). The cosmological parameters ${\rm H_0=70 \,
km~s^{-1}~Mpc^{-1}}$, $\Omega_{\rm m}$=0.3, $\Omega_{\Lambda}$ = 0.7
are used throughout the paper, and the spectral index $\alpha$ is
defined as $f_{\nu}$ $\propto$ $\nu^{-\alpha}$ with $f_{\nu}$ being
the flux density at frequency $\nu$.

\section{Sample and data analysis}

The parent sample of this work is 185 FSRQs in Chen et al. (2009).
It was constructed through cross-correlating the Shen et al. (2006)
SDSS DR3 X-ray quasar sample (3366 sources, see Shen et al. 2006 for
details) with Faint Images of the Radio Sky at Twenty-Centimeters
1.4 GHz radio catalogue (FIRST, Becker, White \& Helfand 1995) and
the Green Bank 6 cm radio survey at 4.85 GHz radio catalogue (GB6,
Gregory et al. 1996). The sample of 185 FSRQs was constructed from
conventional definition of FSRQs with a spectral index between 1.4
and 4.85 GHz $\alpha<0.5$. In this work, we select 87 FSRQs with
redshift $z<0.83$ out of 185 FSRQs. The redshift restriction is
selected in order to measure [O III] $\rm 5007\AA$ line from SDSS
spectra as well as $\rm H\beta$.

The SDSS spectra cover the wavelength range from 3800 to 9200 $\rm
\AA$ with a resolution of about 1800 - 2000 (see Schneider et al.
2005 for details). The spectral analysis is briefly described in
this work, and more details can be found in Chen et al. (2009). The
SDSS spectra were firstly corrected for the Galactic extinction
using the reddening map of Schlegel, Finkbeiner \& Davis (1998) and
then shifted to their rest wavelength. We choose those wavelength
ranges as pseudo-continua, which are not affected by prominent
emission lines, and then decompose the spectra into the following
three components: (1) A power-law continuum to describe the emission
from the active nucleus. (2) An Fe II template adopting the UV Fe II
template from Vestergaard \& Wilkes (2001), and optical one from
V\'{e}ron-Cetty et al. (2004). (3) A Balmer continuum generated in
the same way as Dietrich et al. (2002).

The modeling of above three components is performed by minimizing
the $\chi^2$ in the fitting process. The final multicomponent fit is
then subtracted from the observed spectrum. The broad emission lines
were measured from the continuum subtracted spectra. For the
redshift range of our sources, we focused on several prominent
emission lines, i.e. $\rm H\alpha$, $\rm H\beta$ and Mg II.
Generally, two gaussian components were adopted to fit each of these
lines, indicating the broad and narrow line components,
respectively. The blended narrow lines, e.g. [O III]
$\lambda\lambda4959,5007\rm \AA$ and [He II] $\rm \lambda 4686 \AA$
blending with $\rm H\beta$, and [S II] $\lambda\lambda6716,6730\rm
\AA$, [N II] $\lambda\lambda6548,6583\rm \AA$ and [O I]$\rm \lambda
6300 \AA$ blending with $\rm H\alpha$, were included as one gaussian
component for each line at the fixed line wavelength. The $\chi^{2}$
minimization method was used in fits. The line width FWHM, line flux
of broad $\rm H\alpha$, $\rm H\beta$, Mg II and narrow [O III]$\rm
5007\AA$ lines were obtained from the final fits for our sample.

There are various empirical relations between the radius of broad
line region (BLR) and the continuum (and/or broad line) luminosity,
which can be used to calculate the black hole mass in combination
with the line width FWHM of broad emission lines. Since the
continuum luminosity of FSRQs are usually contaminated by the
non-thermal jet emission, we use the Vestergaard \& Peterson (2006)
method to calculate $M_{\rm BH}$, which utilizes the FWHM and
luminosity of broad $\rm H\beta$:
\begin{equation}
M_{\rm BH}= 4.68\times10^6 
   \left( \frac{\it L(\rm H\beta)}{10^{42}\, \rm erg~s^{-1}}\right)^{0.63}
  \left(\frac{\rm FWHM(H\beta)}{\rm 1000~km~s^{-1}} \right)^2 {\rm M_{\odot}}
\label{MBH_LHb.eq}
\end{equation}

In this work, we also calculate the BLR luminosity $L_{\rm BLR}$
following Celotti, Padovani \& Ghisellini (1997) by scaling the
strong broad emission lines $\rm H\alpha, \rm H\beta$ and Mg II to
the quasar template spectrum of Francis et al. (1991), in which
Ly$\alpha$ is used as a reference of 100. By adding the contribution
of $\rm H\alpha$ with a value of 77, the total relative BLR flux is
555.77, of which $\rm H\beta$ 22 and Mg II 34 (Celotti et al. 1997;
Francis et al. 1991). 

As shown in Chen et al. (2009), the spectral energy distribution
(SEDs) were constructed for each source using the radio FIRST 1.4
GHz and GB6 4.85 GHz data, the optical data selected as the
line-free spectral region from SDSS spectra, and 1keV X-ray data
compiled in Shen et al. (2006) from ROSAT All Sky Survey. When
available, the 2MASS IR ($J, H,$ and $Ks$) (Skrutskie et al. 2006)
and the Far- and near-UV $GALEX$ data (Martin et al. 2005) are also
added. The synchrotron peak frequency and the corresponding peak
luminosity were thus obtained for each source through fitting the
SED with a third-degree polynomial following Fossati et al. (1998).
The thermal emission from accretion disk and host galaxy is
estimated in Chen et al. (2009). The accretion disk emission is
calculated assuming a steady geometrically thin, optically thick
accretion disk around a Schwarzschild black hole and using estimated
black hole mass and bolometric luminosity. The host galaxy spectra
is estimated using the elliptical galaxy template of Mannucci et al.
(2001), in combination with the bulge absolute luminosity in R-band
estimated from $M_{\rm BH} - M_{\rm R}$ relation of McLure et al.
(2004). The total contribution of accretion disk and host galaxy
thermal emission are estimated by calculating the fraction of the
thermal emission to the SED data at SDSS optical and $GALEX$ UV
region. A marginal value of 50\% at most of SED wavebands is used to
divide the FSRQs into thermal-dominated ($>50\%$) and
nonthermal-dominated ($<50\%$). Using this criterion, 36 of 87 FSRQs
in this work are non-thermal dominant FSRQs, and 51 are
thermal-dominated ones. The [O III] measurements are finally
performed in 81 FSRQs, and [O III] line is rather weak in remaining
6 FSRQs, which therfore can not be properly measured. Among 81 [O
III] available FSRQs, 32 sources are non-thermal dominant FSRQs, and
49 are thermal-dominated FSRQs.

\section{results and discussion}

\subsection{$M_{\rm bh}$ - $\rm \sigma_{\rm [O III]}$ relation}

The relationship between the black hole mass $M_{\rm bh}$ and
velocity dispersion $\rm \sigma_{\rm [O III]}=\rm FWHM ({\rm [O III]
5007\AA})/2.35$ for all 81 FSRQs is shown in Fig. 1, in which the
Tremaine et al. relation of equation (1) is also plotted as dashed
line. 
From the figure, the significant scatter and deviation from Tremaine
et al. relation is apparently seen, with majority of sources lying
above the relation, which implies a higher black hole mass for our
FSRQs than nearby inactive galaxies at given velocity dispersion.
This result is actually consistent with Liu \& Jiang (2006) and Bian
et al. (2008) for general radio-loud AGNs/quasars. Both works
claimed that radio-quiet AGNs evenly follow the Tremaine et al.
$M_{\rm bh} - \sigma_{*}$ relation, however, radio-loud ones deviate
with the same trend as ours. Alternatively, radio-loud quasars are
found to have relatively smaller [O III] line width than radio quiet
quasars at given black hole mass or bulge luminosity (e.g. Shields
et al. 2003; Bonning et al. 2005). However, the reason is still
unknown.

\begin{figure}
\begin{center}
\includegraphics[width=0.4\textwidth]{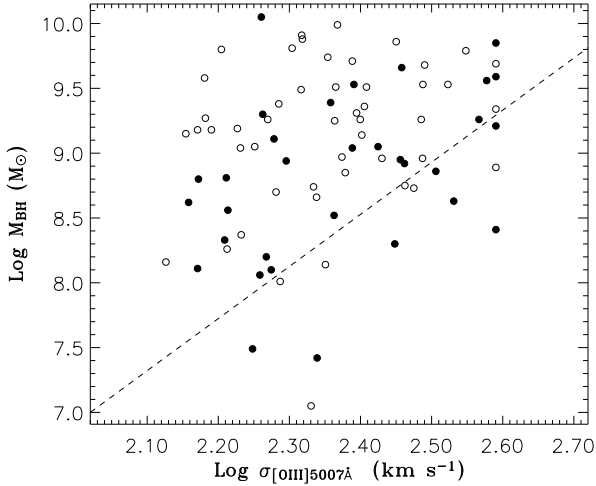}
\caption{The velocity dispersion $\rm \sigma_{[O III]}$ versus black
hole mass. The dashed line is the $M_{\rm bh} - \sigma_{*}$ relation
of Tremaine et al. (2002) for 31 nearby inactive galaxies. The solid
circles are non-thermal dominant FSRQs, while the open circles are
thermal-dominant FSRQs.} \label{fig1}
\end{center}
\end{figure}

The relation between the BLR radius and $\rm H\beta$ luminosity, and
the corresponding empirical relation to estimate black hole mass in
equation (2), are basically scaled from the AGNs sample with
reverberation mapping black hole mass, of which most of sources are
radio-quiet. The main advantage of equation (2) in estimating black
hole mass for FSRQs is that the contamination of non-thermal jet
emission (likely Doppler boosted) in continuum luminosity can be
avoided. However, it is not proved yet whether it is applicable to
radio-loud AGNs. As an evaluation of jet emission contamination, Liu
et al. (2006) shown that the optical continuum luminosity of FSRQs
usually exceed the value estimated from the $L_{\rm H\beta} - L_{\rm
5100\AA}$ relation $L_{\rm 5100\AA}=0.843\times10^{2}L_{\rm
H\beta}^{0.998}$ fitted for Kaspi et al. (2000) radio quiet AGNs
using ordinary least-square (OLS) linear fit method (see Fig. 4 in
Liu et al. 2006). This result is consistent with the expectation
that the continuum emission of FSRQs are contaminated by the (even
dominant) non-thermal jet emission. As a check for our FSRQs sample,
we plot the relationship between $L_{\rm H\beta}$ and $L_{\rm
5100\AA}$ in Fig. 2. We found that the thermal FSRQs well follow the
$L_{\rm H\beta} - L_{\rm 5100\AA}$ relation for radio quiet AGNs of
Liu et al. (2006), supporting the dominance of thermal emission in
these 51 FSRQs.
However, large scatter exist for 36 nonthermal FSRQs.
While most sources follow the relation, about 10 of 36 sources
deviate from the relation, with larger luminosity at $\rm 5100\AA$
than expectations from relation, which is most likely due to the
contribution of non-thermal jet emission.

\begin{figure}
\begin{center}
\includegraphics[width=0.4\textwidth]{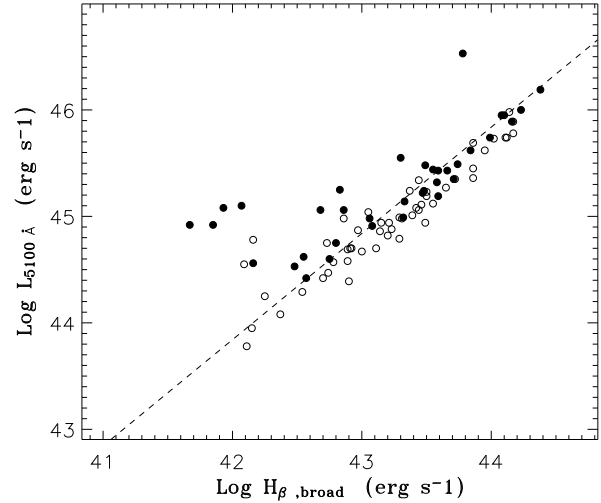}
\caption{The luminosity at $\rm 5100\AA$ versus broad $\rm H\beta$
luminosity. The symbols are same as in Fig. 1. The dashed line is
the OLS bisector linear fit to Kaspi et al. (2000) radio quiet AGNs
(see Liu et al. 2006).} \label{fig2}
\end{center}
\end{figure}

The consistence of $L_{\rm H\beta} - L_{\rm 5100\AA}$ relation of
the thermal-dominant FSRQs with that of radio quiet AGNs implies
that the thermal continuum emission and the broad line emission
follow the similar relation. Therefore, there will be likely little
problem in estimating the BLR radius using empirical $L_{\rm H\beta}
- R_{\rm BLR}$ relation, unless the BLR kinematics (e.g. BLR radius)
in our FSRQs is largely different from that of radio quiet AGNs.
However, it is rather difficult to move our sources downward to
follow the Tremaine et al. relation solely by changing BLR radius,
let alone to reduce the large scatter in Fig. 1. For radio-loud
quasars, especially FSRQs, the disk-like BLR geometry (Wills \&
Browne 1986) can cause the underestimate of black hole mass (Lacy et
al. 2001; McLure \& Dunlop 2001). However, this effect will even
make situation more sever, i.e. our FSRQs will deviate towards
larger black hole mass from the Tremaine et al. relation. The likely
explanation of smaller [O III] line width in radio-loud quasars can
be a different [O III] kinematics or geometry in radio-loud quasars,
especially for our FSRQs. However, the reason is unclear. The clues,
if any, may be from the recent finding that the [O III] luminosity
is not emitted isotropically and that there is significant
extinction towards or within the narrow line region in a subset of
Seyfert 2 galaxies, in the way that the observed [O III] luminosity
are systematically smaller for obscured Seyferts by comparing [O
III] with [O IV] 25.9 $\rm \mu m$ for a unbiased Seyfert galaxies
sample (Diamond-Stanic et al. 2009; see also Jackson \& Browne 1990;
Hass et al. 2005; Mel\'{e}ndez et al. 2008; Zhang et al. 2008). More
relevant to our FSRQs as radio powerful sources, the [O III]
emission from quasars was compared with radio galaxies by Jackson \&
Browne (1990), and they found that the [O III] emission of quasars
is much stronger than that of radio galaxies, which was interpreted
as part of the [O III] emission being obscured by the torus in radio
galaxies. The similar conclusion was reached by Haas et al. (2005)
by comparing [O III] with [O IV] 25.9 $\rm \mu m$ for a sample of
seven 3CR FR II radio galaxies and seven 3CR quasars. It is well
established that the NLR is stratified with high-density and high
ionization gas at close to the continuum source whereas low density
and low-ionization gas is in the outer part of the NLR (Nagao et al.
2003; Riffel et al. 2006). Zhang et al. (2008) claimed that a
significant fraction of the [O III] emission, at least for their
Seyfert galaxies, may arise from the inner dense part of the NLR,
which however can be covered thus obscured by the torus with its
inner edge on scales of parsecs, and its extent likely on scales of
several tens of parsecs (e.g.,
Schmitt et al. 2003, Jaffe et al. 2004). 
Due to the complex of [O III] line profile (Bonning et al. 2005),
nevertheless, it needs further investigation on the $M_{\rm bh} -
\sigma_{*}$ relation for our FSRQs using other narrow lines, e.g. [S
II] (Greene \& Ho 2005). However, the redshift range will be much
limited.


Although the detailed reason is unclear, the deviation of our FSRQs
in Fig. 1 could be related with several recent findings on the bulge
- black hole relations (e.g., Aller \& Richstone 2007; Hopkins et
al. 2007; Lauer et al. 2007). The black hole masses predicted from
the $M_{\rm bh} - \sigma$ relation are found to be in conflict with
those from the $M_{\rm bh} - L_{\rm bulge}$ relation for
high-luminosity galaxies (e.g. brightest cluster galaxies: BCGs),
with the former relation predicting a larger black hole mass (Lauer
et al. 2007). While this may be explained by the slow increase in
$\sigma$ with $L_{\rm bulge}$ and the more rapid increase in
effective radii with $L_{\rm bulge}$ seen in BCGs as compared to
less luminous galaxies, the authors argued that the $M_{\rm bh} -
L_{\rm bulge}$ relationship is a plausible description for galaxies
of high luminosity. From the major galaxy merger simulations,
Hopkins et al. (2007) found the so-called black hole mass
fundamental plane in the form of $M_{\rm bh}\propto
M_{*}^{0.54\pm0.17}\sigma^{2.2\pm0.5}$, similar to relations found
observationally, where $M_{*}$ is the stellar mass. They also
claimed that this fundamental plane is better than any
single-variate predictor of black hole mass, e.g. the $M_{\rm bh} -
\sigma$ relation and the $M_{\rm bh} - M_{*}$ relation. Moreover,
the bulge gravitational binding energy $E_{\rm g}$ may play a
dominant role, in that $M_{\rm bh}\propto E_{\rm g}^{0.6}$ as found
by Aller \& Richstone (2007), which is as strong a predictor of
$M_{\rm bh}$ as the velocity dispersion $\sigma$, for the elliptical
galaxies. In view of these findings, it is possible that our FSRQs,
especially those with high black hole mass (e.g. $M_{\rm
bh}>3\times10^9\rm M_{\odot}$), may have larger bulge mass at a
given $\sigma$. Alternatively, if these objects lie in the center of
galaxies that continue to accrete the surrounding ICM after the
initial black hole growth, as in so-called radio-mode feedback
models (e.g., Croton et al., 2006), then perhaps the black hole can
grow to a higher mass than the mass when it originally settled on
the $M_{\rm bh} - \sigma$ relationship. As shown in Croton et al.
(2006) (see their Fig. 3.), the growth of black holes is dominated
by the `quasar mode' at high redshift and falls off sharply at
$z\lesssim 2$, in contrast, the `radio mode' becomes important at
low redshifts where it suppresses cooling flows.

The deviation of $M_{\rm bh}$ from Tremaine et al. relation can be
calculated as $\Delta M_{\rm bh}=M_{\rm bh}-M_{\rm bh}(\sigma_{\rm
[O III]})$, of which $M_{\rm bh}(\sigma_{\rm [O III]})$ is the
predicted black hole mass from equation (1) using [O III] velocity
dispersion. We perform the correlation analysis between $\Delta
M_{\rm bh}$ and other parameters, i.e. redshift, Eddington ratio
$L_{\rm bol}/L_{\rm Edd}$, 5 GHz luminosity $L_{\rm 5 GHz}$, and
synchrotron peak luminosity $L_{\rm peak}$, and only found a
moderately significant correlation between $\Delta M_{\rm bh}$ and
redshift with Spearman correlation coefficient $r=0.296$ at
confidence level
$\sim 99.3\%$. 
However, this correlation may likely be caused by the common
dependence of other parameters, e.g. $L_{\rm H\beta}$. Indeed, the
partial Spearman correlation analysis shown that the strong
correlation no longer exist independent of $L_{\rm H\beta}$.
Therefore, there is no strong evidence of cosmology evolution of
$M_{\rm bh} - \sigma_{\rm [O III]}$ relation up to $z=0.83$. This is
in contrast to the results of several studies (e.g. Woo et al. 2006;
Shields et al. 2006) that have found apparent positive correlations
between the $M_{\rm bh} - \sigma_{*}$ relation and redshift.
However, Shields et al. (2003) and Salviander et al. (2006), found
no evidence for redshift evolution up to $z\sim3$, using [O III]
line widths as surrogates for $\sigma_{*}$.

\subsection{Synchrotron peak luminosity, broad and narrow line region luminosity}

The broad and narrow emission lines are believed to be produced from
photonionization process, although the shock-excitation from jet-ISM
interaction can not be ignored for narrow emission lines in some
AGNs (e.g. Dopita \& Sutherland 1995). Therefore, the intimate
relation is expected between broad and narrow line luminosities.
Indeed, a significant correlation between the [O III] luminosity and
BLR one is found with Spearman correlation coefficient $r=0.875$ at
confidence level of $\gg99.99\%$. Using partial Spearman correlation
analysis to exclude the common dependence of both luminosities on
redshift, the correlation is still significant with correlation
coefficient $r=0.664$ at confidence level of $\gg99.99\%$.

\begin{figure}\label{fig3}
\begin{center}
\includegraphics[width=0.4\textwidth]{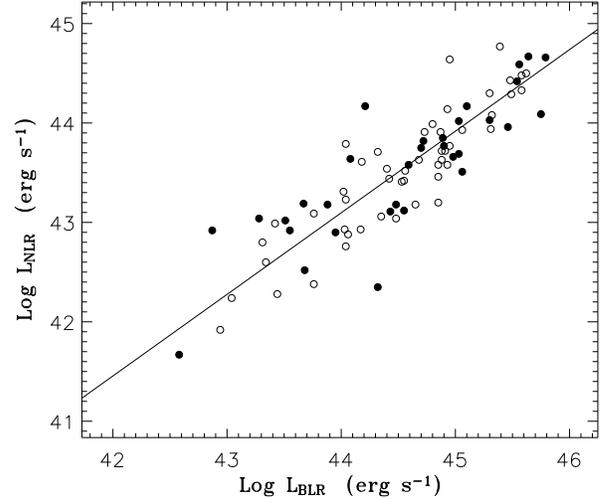}
\caption{The BLR luminosity versus NLR luminosity. The symbols are
same as in Fig. 1. The solid line is the OLS bisector linear fit for
all 81 FSRQs.}
\end{center}
\end{figure}

In this work, we tentatively transfer the [O III] $\rm 5007\AA$
luminosity to NLR luminosity, using relation of $L_{\rm
NLR}=3\times(3\times L_{\rm [O II]}+1.5\times L_{\rm [O III]})$ and
assuming $L_{\rm [O III]}=4\times L_{\rm [O II]}$ (Rawlings \&
Saunders 1991). The relationship between the luminosity of BLR and
NLR is presented in Fig. 3. The ordinary least-square (OLS) bisector
linear fit to $L_{\rm BLR}$ - $L_{\rm NLR}$ relation gives:
\begin{equation} \rm log~\it L_{\rm NLR}\rm =(0.89\pm0.05)~\rm
log~\it L_{\rm BLR}\rm +(3.91\pm2.10)
\end{equation}
The mean value of log $L_{\rm NLR}$ is $\rm \langle log~\it L_{\rm
NLR} \rangle \rm =43.52\pm0.66$, and for log $L_{\rm BLR}$ is $\rm
\langle log~\it L_{\rm BLR} \rangle \rm =44.53\pm0.75$. The BLR
luminosity is, on average, larger than NLR luminosity by about one
order of magnitude with $\rm \langle log~\it (L_{\rm BLR}/L_{\rm
NLR}) \rangle \rm =1.01\pm0.37$. This result means that the covering
factor of NLR is about one tenth of that of BLR. Moreover, it is
consistent within a factor of three with the ratio of BLR to NLR
luminosities ($\sim25$) calculated by using the relative flux in the
composite quasar spectrum of Francis et al. (1991), which shows that
our FSRQs spectra generally follow the composite
spectrum although they are radio powerful sources. 
In addition, the scaling between the BLR and NLR luminosities
enables us to statistically estimate the bolometric luminosity from
narrow emission lines, which can be more readily observed, e.g. for
type 2 AGNs. If using $L_{\rm bol}=30L_{\rm BLR}$ newly calibrated
in Xu, Cao \& Wu (2009), statistically, the bolometric luminosity
can be estimated as $L_{\rm bol}=300L_{\rm NLR}$, i.e. $L_{\rm
bol}=2025 ~L_{\rm [O III]}$, implying a covering factor of $\sim
0.003$ for narrow line region, similar to those of Rawlings \&
Saunders (1991) ($\sim0.01$) and Willott et al. (1999)
($\sim0.003$). More specifically, the calibration between the
bolometric luminosity and NLR luminosity can be given through OLS
bisector linear fit:
\begin{equation} \rm log~\it L_{\rm bol}\rm =(1.12\pm0.06)~\rm
log~\it L_{\rm NLR}\rm -(3.92\pm2.59)
\end{equation}
Our bolometric correction is consistent with Heckman et al. (2004)
($L_{\rm bol}/L\rm ([O III])\approx3500$ for Seyfert galaxies with a
scatter of approximately 0.38 dex) and Willott et al. (1999)
($L_{\rm bol}\simeq5\times10^3L_{\rm [O II]}$ W). The consistent
results are also recently found by Wu (2009b) that $L_{\rm
bol}\propto L_{\rm [O III]}^{0.95\pm0.08}$ for a sample of Seyfert 1
galaxies and radio-quiet quasars using $L_{\rm bol}=9\lambda
L_{\lambda}(\rm 5100\AA)$, and the mean bolometric correction
$L_{\rm bol}/L\rm ([O III])\sim3400$, 2000 for 23 radio-quiet
quasars and 20 Seyfert 1 galaxies, respectively.

The relationship between the NLR luminosity and synchrotron peak
luminosity is given for all 81 FSRQs in Fig. 4. We found a strong
correlation between two parameters with a Spearman correlation
coefficient $r=0.735$ at confidence level $\gg 99.99\%$. The partial
correlation analysis gives a significant correlation with
correlation coefficient $r=0.408$ at confidence level $\sim 99.98\%$
independent of redshift. When only 32 non-thermal dominant FSRQs are
considered, the original significant correlation ($r=0.637$ at
confidence level $\sim 99.96\%$) still remains ($r=0.406$ at
confidence level $\sim 98\%$). For these 32 non-thermal dominant
FSRQs, the OLS bisector linear fit on $\nu L_{\rm \nu peak}$ -
$L_{\rm NLR}$ relation gives:
\begin{equation}
\rm log~\it \nu L_{\rm \nu peak}\rm =(0.80\pm0.10)~\rm log~\it
L_{\rm NLR}\rm +(10.46\pm4.32)
\end{equation}
which remains same when fitting on all 81 FSRQs. The strong
correlation is also found between the NLR and 5 GHz luminosity (see
Fig. 4), however, it is less significant ($r=0.463$ at confidence
level $\sim 99.99\%$) than that between the NLR luminosity and
synchrotron peak luminosity.

\begin{figure}
\begin{center}
\includegraphics[width=0.4\textwidth]{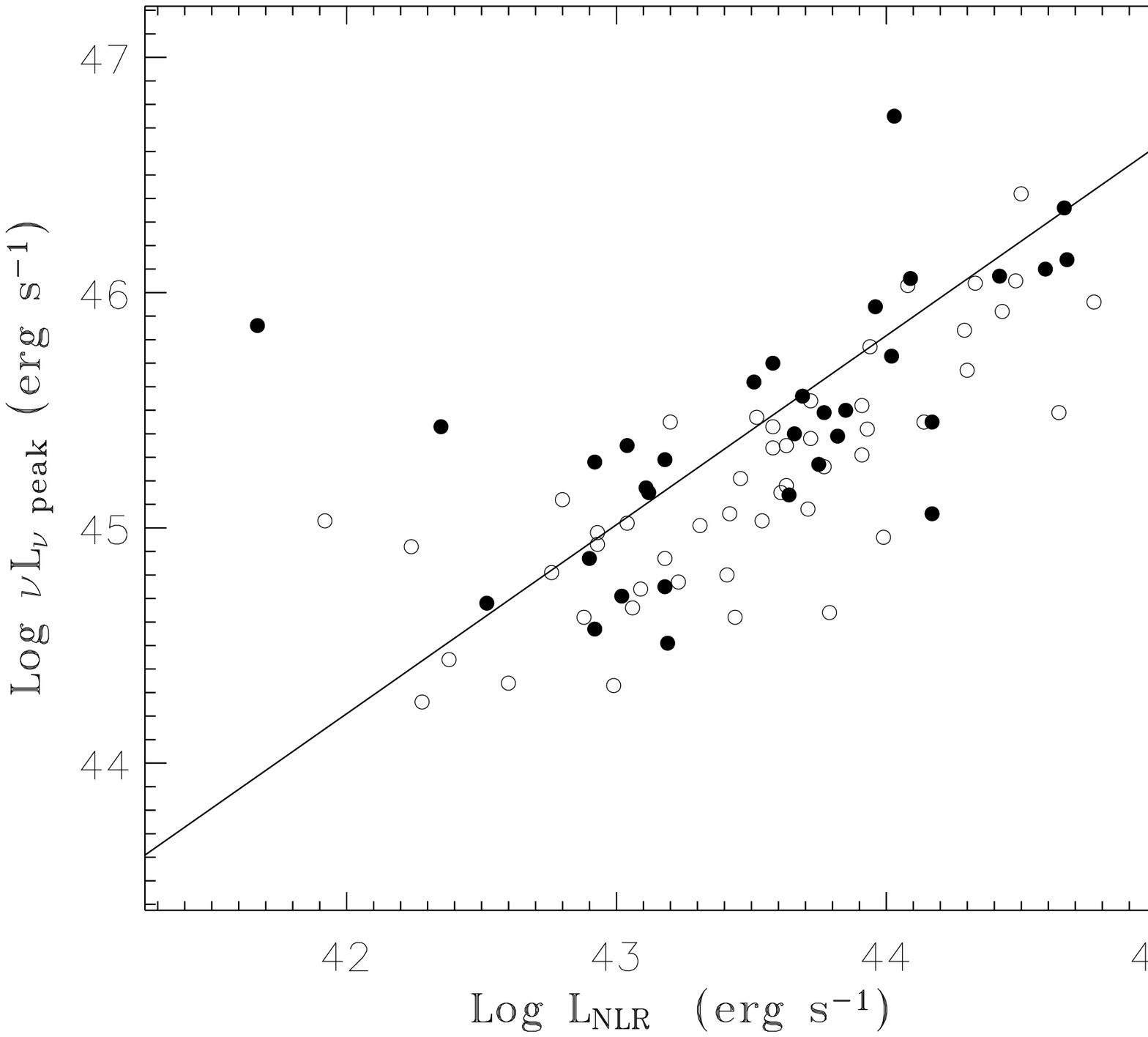}
{\vspace{0.1cm}}
\includegraphics[width=0.4\textwidth]{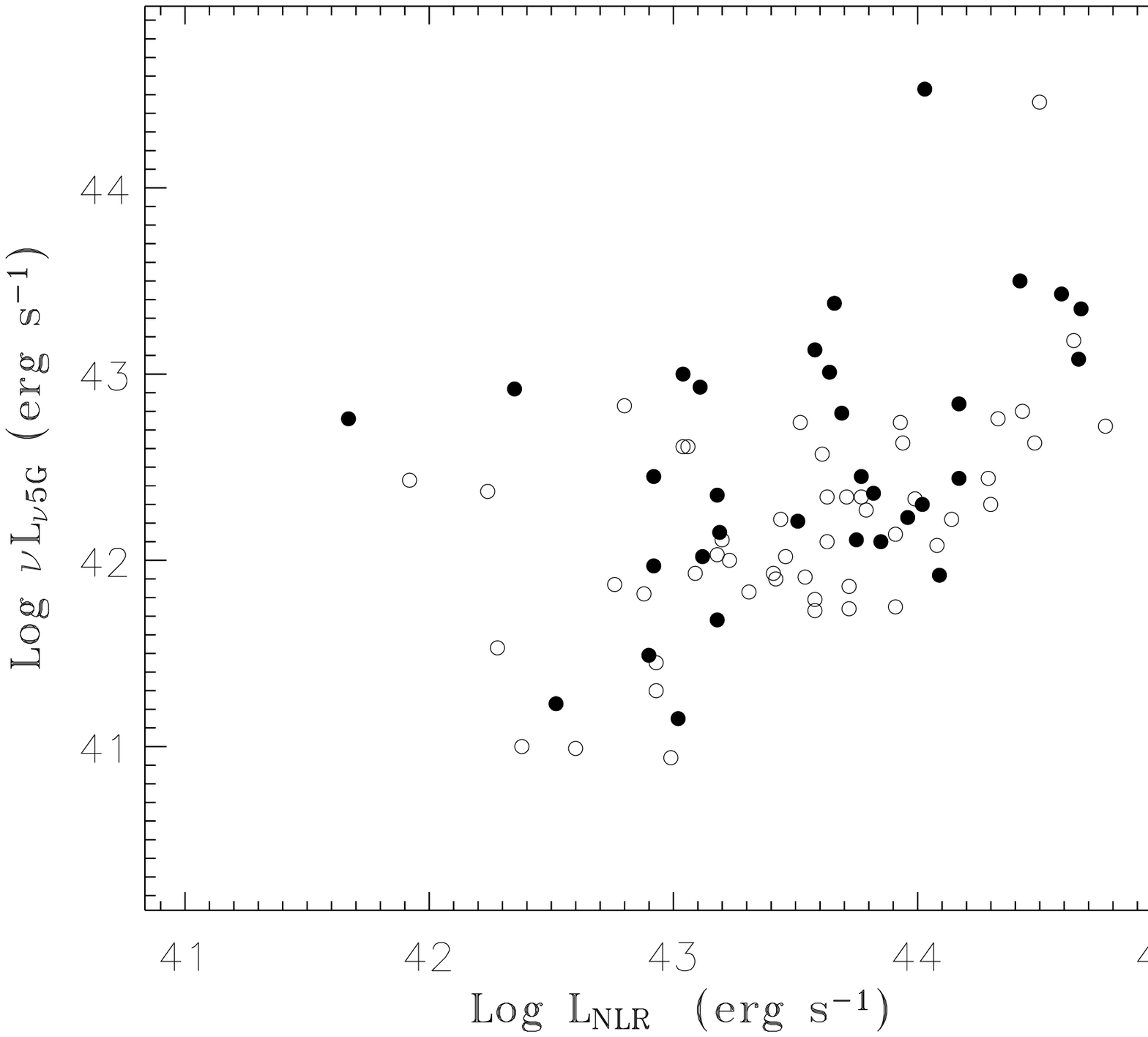}
\caption{The upper panel: The synchrotron peak luminosity versus NLR
luminosity. The symbols are same as in Fig. 1. The solid line is the
OLS bisector linear fit for 32 non-thermal dominant FSRQs. The lower
panel: 5 GHz luminosity versus NLR luminosity. The symbols are same
as in Fig. 1.}
\end{center}
\label{fig4}
\end{figure}

The advantages of using the synchrotron peak luminosity is that the
most of synchrotron emission are radiated at synchrotron peak
frequency, at which the luminosity can be a good indicator of
synchrotron emission. Since the synchrotron peak frequency varies
from source to source, the luminosity at fixed waveband is actually
from the different portion of source SED. This can be strengthened
from Fig. 4, in which the large scatter in $\nu L_{\nu\rm 5
G}-L_{\rm NLR}$ is significantly reduced in $\nu L_{\nu\rm
peak}-L_{\rm NLR}$, and with the less significance of the former
correlation. This not only implies a tight relation between jet
physics and accretion process, but also claims that the synchrotron
peak luminosity can be a better indicator of jet emission than 5 GHz
one. However, the possibility that part, if not all, of the $\nu
L_{\nu\rm peak}-L_{\rm NLR}$ correlation can be caused from
shock-excitation for NLR by the jet-ISM interactions (e.g. Dopita \&
Sutherland 1995; Bicknell et al. 1997), can not be completely
excluded. By comparing the [O IV] 25.9 $\rm \mu m$ and [O III]
luminosities for the combined sample of radio loud and radio quiet
Seyfert galaxies, Mel\'{e}ndez et al. (2008) found that radio loud
sources exhibit higher emission-line luminosities than those of
radio quiet ones. The authors claimed that this result could be
explained by a proposed bow shock model where part of the NLR
emission is being powered by radio-emitting jets, present in
radio-loud AGNs. Moreover, the jet-ISM interaction are invoked to
explain the highly broadened [O III] line profiles in young radio
galaxies (CSS/GPS sources) and some luminous linear radio sources
(e.g., Gelderman \& Whittle 1994; Nelson \& Whittle 1996).

Although it can be a good indicator of the jet emission, the defect
of the synchrotron peak luminosity lies in the contamination from
beaming effect, which precludes it to well indicate the intrinsic
source power. Only when the Doppler boosting is known for each
source, the intrinsic source power can be obtained, which however
can not be performed at present stage.


\section*{Acknowledgements}

We thank the anonymous referee for insightful comments and
constructive suggestions. We thank Qingwen Wu for valuable
discussions. This work is supported by National Science Foundation
of China (grants 10633010, 10703009, 10833002, 10773020 and
10821302), 973 Program (No. 2009CB824800), and the CAS
(KJCX2-YW-T03).

Funding for the SDSS and SDSS-II has been provided by the Alfred P.
Sloan Foundation, the Participating Institutions, the National
Science Foundation, the U.S. Department of Energy, the National
Aeronautics and Space Administration, the Japanese Monbukagakusho,
the Max Planck Society, and the Higher Education Funding Council for
England. The SDSS Web Site is http://www.sdss.org/. The SDSS is
managed by the Astrophysical Research Consortium for the
Participating Institutions. The Participating Institutions are the
American Museum of Natural History, Astrophysical Institute Potsdam,
University of Basel, University of Cambridge, Case Western Reserve
University, University of Chicago, Drexel University, Fermilab, the
Institute for Advanced Study, the Japan Participation Group, Johns
Hopkins University, the Joint Institute for Nuclear Astrophysics,
the Kavli Institute for Particle Astrophysics and Cosmology, the
Korean Scientist Group, the Chinese Academy of Sciences (LAMOST),
Los Alamos National Laboratory, the Max-Planck-Institute for
Astronomy (MPIA), the Max-Planck-Institute for Astrophysics (MPA),
New Mexico State University, Ohio State University, University of
Pittsburgh, University of Portsmouth, Princeton University, the
United States Naval Observatory, and the University of Washington.


\begin{thebibliography}{99}

\bibitem[]{} Aller M. C., Richstone D. O., 2007, ApJ, 665, 120
\bibitem[Becker, White \& Helfand(1995)]{bec95} Becker R.~H., White R.~L., Helfand D.~J., 1995, \apj, 450, 559
\bibitem[Bian \& Zhao(2004)]{bia04} Bian W., Zhao Y., 2004, MNRAS, 347, 607
\bibitem[Bian et al.(2008)]{bia08} Bian W. H., Chen Y., Hu C., Huang K., Xu Y., 2008, Chinese J. Astron. Astrophys., 8, 522
\bibitem[]{} Bicknell G. V., Dopita M. A., O'Dea C. P. O., 1997, ApJ, 485, 112
\bibitem[Bonning et al.(2005)]{bon05} Bonning E. W., Shields G. A., Salviander S., McLure R.
J., 2005, ApJ, 626, 89
\bibitem[Celotti, Padovani \& Ghisellini(1997)]{cel97}
Celotti A., Padovani P., Ghisellini G., 1997, \mnras, 286, 415
\bibitem[Chen et al.(2009)]{che09} Chen Z. Y., Gu M. F., X. Cao, 2009, MNRAS, arXiv: 0904.1452
\bibitem[]{} Croton D. J. et al., 2006, MNRAS, 365, 11
\bibitem[]{} Diamond-Stanic A. M., Rieke G. H., Rigby J. R., 2009, ApJ, arXiv: 0904.0250
\bibitem[Dietrich et al.(2002)]{die02} Dietrich M., Appenzeller I., Vestergaard M., Wagner S. J., 2002, \apj, 564, 581
\bibitem[]{} Dopita M. A., Sutherland R. S., 1995, ApJ, 455, 468
\bibitem[Ferrarese \& Merritt]{fer00} Ferrarese L., Merritt D., 2000, ApJ, 539, L9
\bibitem[Ferrarese et al.(2001)]{fer01} Ferrarese L., Pogge R. W., Peterson B. M., Merritt D., Wandel A., Joseph
C. L., 2001, ApJ, 555, L79
\bibitem[Fossati et al.(1998)]{fossati98} Fossati G., Maraschi L., Celotti A., Comastri A., Ghisellini G., 1998, \mnras, 299, 433
\bibitem[Francis et al.(1991)]{Francis91} Francis P.~J., Hewett P.~C., Foltz C.~B., Chaffee F.~H., Weymann R.~J., Morris S.~L., 1991, ApJ, 373, 465
\bibitem[Gebhardt et al.(2000a)]{geb00a} Gebhardt K. et al., 2000a, ApJ, 539, L13
\bibitem[Gebhardt et al.(2000b)]{geb00b} Gebhardt K. et al., 2000b, ApJ, 543, L5
\bibitem[]{} Gelderman R., Whittle M., 1994, ApJS, 91, 491
\bibitem[Gregory et al.(1996)]{gre96} Gregory P.~C., Scott W.~K., Douglas K., Condon J.~J., 1996, \apjs, 103, 427
\bibitem[Greene \& Ho(2005)]{gre05} Greene J. E., Ho L. C., 2005, ApJ, 627, 721
\bibitem[]{} Haas M., Siebenmorgen R., Schulz B., Kr\"{u}gel E., Chini R., 2005, A\&A, 442, L39
\bibitem[Heckman et al.(2004)]{hec04} Heckman T. M., Kauffmann G., Brinchmann J., Charlot S., Tremonti C., White S. D. M., 2004, ApJ, 613, 109
\bibitem[]{} Hopkins P. F., Hernquist L., Cox T. J. Robertson B., Krause E., 2007, ApJ, 669, 45
\bibitem[]{} Jackson N., Browne I. W. A., 1990, Nature, 343, 43
\bibitem[]{} Jaffe A. H. et al., 2004, ApJ, 615, 55
\bibitem[Kaspi et al.(2000)]{kas00} Kaspi S., Smith P. S., Netzer H., Maoz D., Jannuzi B. T., Giveon
U., 2000, ApJ, 533, 631
\bibitem[Kong et al.(2006)]{kon06} Kong M. Z., Wu X. B., Wang R., Han J. L., 2006, Chinese J. Astron. Astrophys., 6,396
\bibitem[Kormendy \& Richstone(1995)]{kor95} Kormendy J., Richstone D., 1995, ARA\&A, 33, 581
\bibitem[Lacy et al.(2001)]{lac01} Lacy M., Laurent-Muehleisen S. A., Ridgway S. E., Becker R. H., White
R. L., 2001, ApJ, 551, L17
\bibitem[]{} Lauer T. R. et al., 2007, ApJ, 662, 808
\bibitem[Liu \& Jiang(2006)]{liu06} Liu Y., Jiang D. R., 2006, Chinese J. Astron. Astrophys., 6, 655
\bibitem[Liu, Jiang \& Gu(2006)]{liu06} Liu Y., Jiang D. R., Gu M. F., 2006, ApJ, 637, 669
\bibitem[Magorrian et al.(1998)]{mag98} Magorrian J. et al., 1998, AJ, 115, 2285
\bibitem[Mannucci et al.(2001)]{man01} Mannucci F., Basile F., Poggianti B. M., Cimatti A., Daddi E., Pozzetti L., Vanzi L, 2001, \mnras, 326, 745
\bibitem[]{} Martin D. C., Fanson J, Schiminovich D., Morrissey P., 2005, ApJ, 619, L1
\bibitem[Mclure \& Dunlop(2001)]{mcl01} McLure R. J., Dunlop J. S., 2001, MNRAS, 327, 199
\bibitem[McLure et al.(2004)]{mcl04} McLure R. J., Willott C.J., Jarvis M. J., Rawlings S., Hill G. J., Mitchell E., Dunlop J. S., Wold M., 2004, \mnras, 351, 347
\bibitem[]{} Mel\'{e}ndez M. et al., 2008, ApJ, 682, 94
\bibitem[]{} Nagao T., Murayama T., Shioya Y., Taniguchi Y., 2003, AJ, 126, 1167
\bibitem[Nelson \& Whittle(1996)]{nel96} Nelson C. H., Whittle M., 1996, ApJ, 465, 96
\bibitem[Nelson(2000)]{nel00} Nelson C. H., 2000, ApJ, 544, L91
\bibitem[Rawlings \& Saunders(1991)]{raw91} Rawlings S., Saunders R., 1991, Nature, 349, 138
\bibitem[]{} Riffel R., Rodr\'{i}guez-Ardila A., Pastoriza M. G., 2006, A\&A, 457, 61
\bibitem[Salviander et al.(2006)]{sal06} Salviander S., Shields G. A., Gebhardt K., Bonning E. W., 2006, New Astron. Rev., 50, 803
\bibitem[Salviander et al.(2007)]{sal07} Salviander S., Shields G. A., Gebhardt K.,
Bonning E. W., 2007, ApJ, 622, 131
\bibitem[Schlegel, Finkbeiner \& Davis(1998)]{sch98} Schlegel D. J., Finkbeiner D. P., Davis M., 1998, ApJ, 500, 525
\bibitem[]{} Schmitt H. R., Donley J. L., Antonucci R. R. J., Hutchings J. B., Kinney A. L., Pringle J. E., 2003, ApJ, 597, 768
\bibitem[Schneider et al.(2005)]{sch05} Schneider D. P. et al., 2005, AJ, 130, 367
\bibitem[Shen et al.(2008)]{she08} Shen J. J., Vanden Berk D. E., Schneider D. P., Hall P. B., 2008, AJ, 135, 928
\bibitem[Shen et al.(2006)]{She06} Shen S. Y., White S. D. M., Mo H. J., Voges W., Kauffmann G., Tremonti C., Anderson S.~F., 2006, \mnras, 369, 1639
\bibitem[Shields et al.(2003)]{shi03} Shields G. A., Gebhardt K., Salviander S., Wills B. J., Xie B.,
Brotherton M. S., Yuan J., Dietrich M., 2003, ApJ, 583, 124
\bibitem[Shields et al.(2006)]{shi06} Shields G. A., Menezes K. L., Massart C. A., Vanden Bout P., 2006, ApJ, 641, 683
\bibitem[]{} Skrutskie M. F. et al., 2006, AJ, 131, 1163
\bibitem[Tremaine et al.(2002)]{tre02} Tremaine S. et al., 2002, ApJ, 574, 740
\bibitem[V\'{e}ron-Cetty et al.(2004)]{ver04} V\'{e}ron-Cetty M. P., Joly M., V\'{e}ron P., 2004, A\&A, 417, 515
\bibitem[Vestergaard \& Wilkes(2001)]{vestergaard01} Vestergaard M.,  Wilkes B.~J., 2001, \apjs, 134, 1
\bibitem[Vestergaard \& Peterson(2006)]{Vestergaard06} Vestergaard M., Peterson B.~M., 2006 \apj, 641, 689
\bibitem[Willott et al.(1999)]{wil99} Willott C. J., Rawlings S., Blundell K. M., Lacy M., 1999, MNRAS, 309,
1017
\bibitem[Wills \& Browne(1986)]{wil86} Wills B. J., Browne I. W. A., 1986, ApJ, 302, 56
\bibitem[Woo et al.(2006)]{woo06} Woo J.-H., Treu T., Malkan M. A., Blandford R. D., 2006, ApJ, 645, 900
\bibitem[Wu(2009)]{wu09} Wu Q. W., 2009a, MNRAS, submitted
\bibitem[Wu(2009)]{wu09} Wu Q. W., 2009b, ApJ, submitted
\bibitem[Xu, Cao \& Wu(2009)]{xu09} Xu Y. D., Cao X., Wu Q. W., 2009, ApJ, 694, L107
\bibitem[]{} Zhang K., Wang T. G., Dong X. B., Lu H. L., 2008, ApJ, 685, L109

\end{thebibliography}
\end{document}